\documentclass[aps,prb,twocolumn,showpacs,
preprintnumbers,amsmath,amssymb]{revtex4}
\usepackage{bbm}
\usepackage{amsfonts}

\usepackage{graphicx}
\usepackage{dcolumn}
\usepackage{bm}
\usepackage{ifthen}
\usepackage{booktabs}

\begin{document}
\title{Electronic structure interpolation via atomic orbitals}
\author{Mohan Chen}
\affiliation{Key Laboratory of Quantum Information, University of
Science and Technology of China, Hefei, 230026, People's Republic of
China}
\author{G-C Guo}
\affiliation{Key Laboratory of Quantum Information, University of
Science and Technology of China, Hefei, 230026, People's Republic of
China}
\author{Lixin He \footnote{corresponding author: helx@ustc.edu.cn} }
\affiliation{Key Laboratory of Quantum Information, University of
Science and Technology
of China, Hefei, 230026, People's Republic of China}
\date{\today }

\begin{abstract}
We present an efficient scheme for accurate electronic structure
interpolations based on the systematically improvable optimized
atomic orbitals. The atomic orbitals are generated by minimizing the
spillage value between the atomic basis calculations and the converged
plane wave basis calculations on some coarse $k$-point grid. They
are then used to calculate the band structure of the full Brillouin
zone using the linear combination of atomic orbitals (LCAO)
algorithms. We find that usually 16 -- 25 orbitals per atom can give
an accuracy of about 10 meV compared to the full {\it ab initio}
calculations. The current scheme has several advantages
over the existing interpolation schemes. The scheme is easy to
implement and robust which works equally well for metallic systems
and systems with complex band structures. Furthermore, the atomic
orbitals have much better transferability than the Shirley's basis
and Wannier functions, which is very useful for the perturbation
calculations.
\end{abstract}


\maketitle

\section{Introduction}

Very often one needs to calculate physical properties that need
very highly dense $k$ points to get accurate results.
For example, Yao et al. \cite{yao04}
used $2\times10^{6}$ $k$ points in order to get a converged
value of intrinsic anomalous Hall conductivity. Direct
calculations of the properties via the first-principles method are
too expensive in these cases. An efficient electronic structure
interpolation method is therefore a powerful tool
to reduce the computational cost while retaining the
accuracy at the {\it ab initio} level.

Maximally Localized Wannier Functions (MLWFs) \cite{marzari97} have
been demonstrated as a powerful tool for interpolating electronic
structures accurately. \cite{xinjie06, yates07} It has been used to
calculate the anomalous Hall conductivity, \cite{yates07} low-field
Hall conductivity, and magnetic circular dichroism spectrum,
\cite{xinjie06} etc. In this approach, the band connectivity, band
crossings and avoided crossings are treated correctly. The
evaluation of the velocity matrix can also be performed analytically.
Because of their local character, the MLWFs serve as a kind of exact
tight-binding basis of minimal sizes. However, in
practice, it is not always easy to generate highly localized Wannier
functions for metallic systems, and systems with complex band
structures.


The electronic structures can also be interpolated by the Bloch
states. Shirley proposed an interpolation scheme in 1996.
\cite{shirley96} The optimal basis set is extracted from the full
{\it ab initio} calculations on a coarse
$k$-point grid. It has been shown that by
using about 35 basis functions per atom, the optimal
basis can reproduce the band structures of Si and LiF within
an accuracy of 10 meV. Recently, Shirley's
scheme has been extended to more general $k$-point
sampling. \cite{prendergast09} The Shirley's method can be used well
for both insulators and metals. However, the Shirley's basis is
delocalized, therefore it can not take the advantages of the
locality as MLWFs. Furthermore, the Shirley's scheme requires
storing a large number of wave functions, which might be problematic
for large systems.

Atomic orbitals have been popular recently for self-consistent {\it
ab initio} calculations. Relatively compact basis size can give quite
satisfactory results compared to the plane wave calculations.
\cite{kenny00, junquera01, soler02, ozaki03} However, to get
extremely high accuracy band structures (errors $\sim$ 2 -- 3 meV), the
small, rigid atomic basis sets are obviously not adequate.
Recently the authors (CGH) developed a method to generate
systematically improvable fully optimized atomic orbitals for {\it
ab initio} calculations. \cite{chen10} The CGH scheme can generate
orbitals with arbitrary angular momentum and
multi-zeta functions in a unified procedure. The
CGH orbitals offer highly accurate and transferable atomic basis
sets, which have been tested for a wide variety of systems,
including semiconductors, oxides, metals, clusters, etc.

We show in this paper that the CGH method can be revised to give an
efficient scheme for electronic structures interpolations. This
scheme keeps the advantages of both MLWFs and Shirley's methods and
gets rid of some of their disadvantages. The atomic orbitals are highly
localized (more localized than MLWFs), and therefore can take the
advantage of the locality. The atomic orbitals are very neat, and
only the one dimensional radial functions are stored. We show that
the scheme works well for both metals and insulators. The highly
localized orbitals can always be generated for metallic systems.
Additionally, the transferbility of local orbitals for interpolation
purposes is explored. Once the atomic orbitals are generated, the
electronic properties can be calculated efficiently by the well
established linear combination of atomic orbitals (LCAO) algorithms,
with the accuracy of {\it ab initio} plane wave calculations.

The rest of the paper is organized as follows.
In Sec.~\ref{sec:methods}, we introduce briefly the method 
we use to do the
band structure interpolation via atomic orbitals.
In Sec.~\ref{sec:results}, we show, via a few examples,
how the present interpolation scheme works for both metallic
and insulating systems, and systems with complex band structures.
We also discuss the transferability of our interpolation bases.
We conclude in Sec.~\ref{sec:conclusion}.

\section{Methods}
\label{sec:methods}

All the calculations performed in this paper are based on the density
functional theory (DFT) within local (spin) density approximation
[L(S)DA]. Although our method is demonstrated for the
norm-conserving \cite{hamann79} pseudopotentials, in principle, it
should also work well for the ultrasoft or PAW pseudopotentials.
Monkhorst-Pack $k$ points \cite{monkhorst76} are used in the
following calculation.

\subsection{Generating local orbitals}
The details of how to construct systematically optimized atomic
basis for {\it ab initio} calculations have been given in Ref.
\onlinecite{chen10}. Here we give a brief review of our earlier
approach to generating local orbitals. The atomic orbitals are
written as radial functions times spherical harmonic functions,
i.e., $\phi_{\mu}({\bf r})=f_{\mu,l}(r)Y_{lm}(\hat{r})$. The radial
function is taken as the linear combination of Spherical Bessel
functions within certain range $r_c$, i.e.,
\begin{equation}
f_{\mu,l}(r)=\left\{
\begin{array}{ll}
\sum_{q}c_{\mu q}j_l(qr), & r < r_c\\
0 & r \geq r_c \, .\\
\end{array}
\right.
\end{equation}
where $\mu = \{\alpha, i, \zeta, l, m\}$, and $\alpha, i, \zeta, l, m$
are the element type, the index of atom of each element type, the
multiplicity of the radial functions, the angular momentum and the
magnetic quantum number, respectively. $j_l(qr)$ is the spherical
Bessel function, where $q$ satisfy $j_{l}(qr_c)$=0.
The coefficients $c_{\mu q}$ are chosen to minimize the ``spillage'',
\cite{portal95,portal96}
\begin{equation}
\mathcal{S}=\frac{1}{N_{n}N_{k}}\sum_{n=1}^{N_{n}N_{k}}\langle
\Psi_{n}({\bf k}) | 1 - \hat{P}({\bf k}) | \Psi_{n}({\bf k}) \rangle
\, ,
\label{eq:spillage}
\end{equation}
between the Hilbert space spanned by the atomic orbitals and the plane wave
calculations, for selected reference states.
In Eq. \ref{eq:spillage}, $\Psi_{n}({\bf k})$ is the eigenstate of plane wave
calculations, and $N_{n}$ and $N_{k}$ are the number of states of
interest and the number of k-points in the Brillouin zone.
$\hat{P}({\bf k})$ is a projector spanned by the atomic orbitals, i.e.,
\begin{equation}
\hat{P}({\bf k})=\sum_{\mu\nu}|\phi_{\mu}({\bf k})\rangle
S_{\mu\nu}^{-1}({\bf k}) \langle \phi_{\nu}({\bf k})| \, ,
\end{equation}
with
\begin{equation}
\langle {\bf r}|\phi_{\mu}({\bf k})\rangle=\sum_{{\bf
R}}\phi_{\mu}({\bf r} -{\bf r}_{\mu}-{\bf R})e^{i{\bf k}\cdot({\bf
r}_{\mu}+{\bf R})} \, ,
\end{equation}
where $\phi_{\mu}({\bf r})$ is atomic orbitals. ${\bf r}_{\mu}$ is the
atom center of the $\mu$-th orbital, and ${\bf R}$ is the lattice
vector. $S_{\mu\nu}({\bf k})=\langle\phi_{\mu}({\bf
k})|\phi_{\nu}({\bf k})\rangle$ is the overlap matrix element.
The minimization of the spillage is achieved by using a simulated
annealing method.
Because the orbitals generated from this step usually have unphysical
oscillations, an additional step is added to smooth out the
orbitals by minimizing the kinetic energy of each orbital.\cite{chen10}

We have performed extensive tests of this scheme for a wide variety
of systems, including semiconductors, oxides, metals, and clusters, etc.
The results show that the obtained atomic bases are very
satisfactory in both accuracy and transferability.\cite{chen10}
However, to make high accuracy electronic structure interpolation, we
need to modify the scheme for the task.

In Ref. \onlinecite{chen10}, where the aim is to do self-consistent {\it
ab initio} calculations in complex chemical environments, to get
maximal transferability of the basis sets, a set of dimers with
different bond lengths are selected as the reference systems. The
valence states are chosen as the reference states. However, in order
to interpolate the electronic structures with extremely high accuracy
for a particular system, the system itself is used as the reference
system. In many cases, the interested states lie only in a
given energy window (i.e., valence bands as well as conduction
bands), therefore, we can choose these states (instead of all the
valence bands) as the references states.

In the self-consistent LCAO calculations, which are usually applied
to deal with extremely large systems, it is important to keep the
basis size compact. Often used basis are DZP (double zeta orbitals
plus a polar orbital) etc. The small basis set is not enough for
high accuracy electronic structure interpolations. One must use much
higher angular momentum and more zeta orbitals. In these cases, the
CGH scheme has obvious advantages, because it can
generate the fully {\it optimized} orbitals with
arbitrary angular momentum and any number of radial functions for a
given angular momentum.

It has been shown in Ref. \onlinecite{chen10} that increasing the
orbital radius cutoff $r_c$ can reduce the spillage value, and
therefore improve the quality of the orbitals. However, a larger
$r_c$ also means more computational costs.
In the self-consistent calculations, one has to
balance the computational costs and the accuracy, and choose a modest
$r_c$. Yet, for band structure interpolations, the systems
under study are usually considerably
smaller, a larger $r_c$ can be used.

Once we obtain the atomic orbitals, the electronic structures can be
calculated efficiently by using well established LCAO methods.

\section{RESULTS AND DISCUSSIONS}
\label{sec:results}

In this Section, we show via a few examples how the present band
structure interpolation scheme works.
We start the tests from simple solid structures like Na, Si and Al. Na
and Al are metallic while Si is a semiconductor. More complex
materials, such as iron based superconductor BaFe$_{2}$As$_{2}$ are then
tested. We further show that the atomic bases have much better
transferability than the the Bloch bases and
Wannier bases for the purpose of band structure interpolation.

\begin{figure}
\centering
\includegraphics[width=2in]{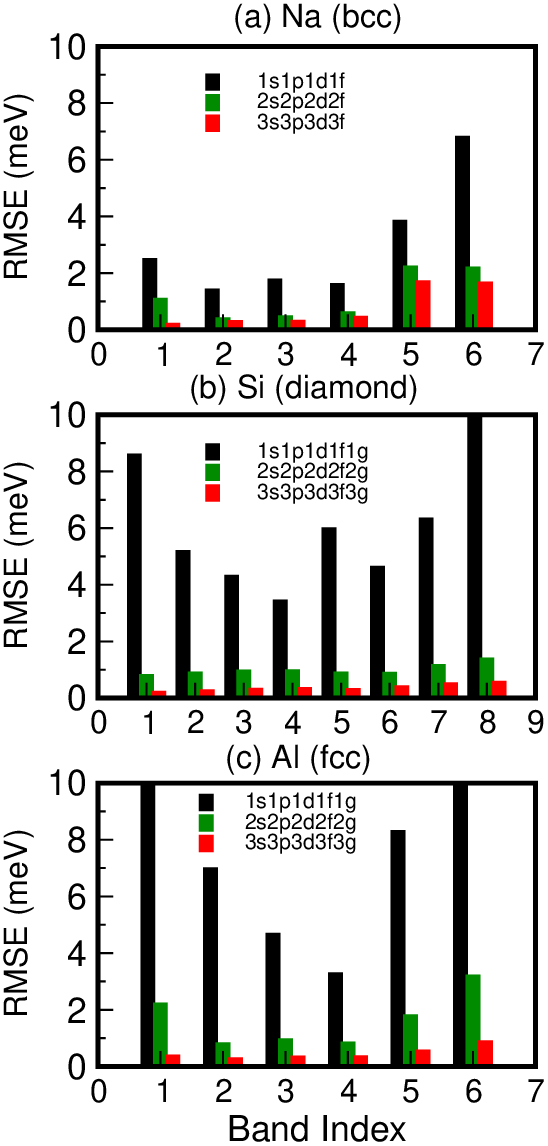}
\caption{(Color online) The RMSE of interpolated bands of (a) bcc
Na; (b) Si diamond structure; (c) fcc Al. The black, red, and green
  columns represent the RMSE of Level 1, 2, 3 atomic basis sets respectively.
}
\label{fig:Si_Na_Al}
\end{figure}

\subsection{Simple Solids}

We take bcc Sodium as the first example.
We use energy cut-off $E_c$=50 Ry. A $8\times8\times8$ $k$-point
mesh is used as the reference states to generate the local orbitals.
We choose the lowest 6 bands for interpolation. In order to quantify
the quality of the interpolation, we calculate the root-mean-square
errors (RMSE) between the interpolated bands and the plane wave
calculations, defined as,
\begin{equation}
{\rm RMSE}= \sqrt{\sum_{i=1}^{N}(\epsilon^{\rm
PW}_{i}-\epsilon^{\rm LCAO}_{i})^{2}/N}\, ,
\end{equation}
where $N$ is the number of the $k$ points that the
comparison has been taken between $\epsilon^{\rm PW}$ the
eigenvalue calculated from plane wave basis and $\epsilon^{\rm
LCAO}$ the eigenvalue calculated from atomic orbitals.

The RMSE of the 6 interpolated bands of Na are shown
in Fig.~\ref{fig:Si_Na_Al}(a). Three levels of atomic orbitals are
generated and compared. At each level, we use one group of s, p, d,
and f orbitals (1s1p1d1f). The radius cutoff $r_c$ is chosen as 12
a.u.  We find that the quality of the basis set is much better if
all the orbitals of the same level are generated simultaneously than
those generated separately.
As shown in Fig.~\ref{fig:Si_Na_Al}(a), the RMSE of the 1s1p1d1f,
2s2p2f2f and 3s3p3d3f local basis sets are less than 8
meV, 3 meV and 2 meV respectively.

\begin{figure}
\centering
\includegraphics[width=3.2in]{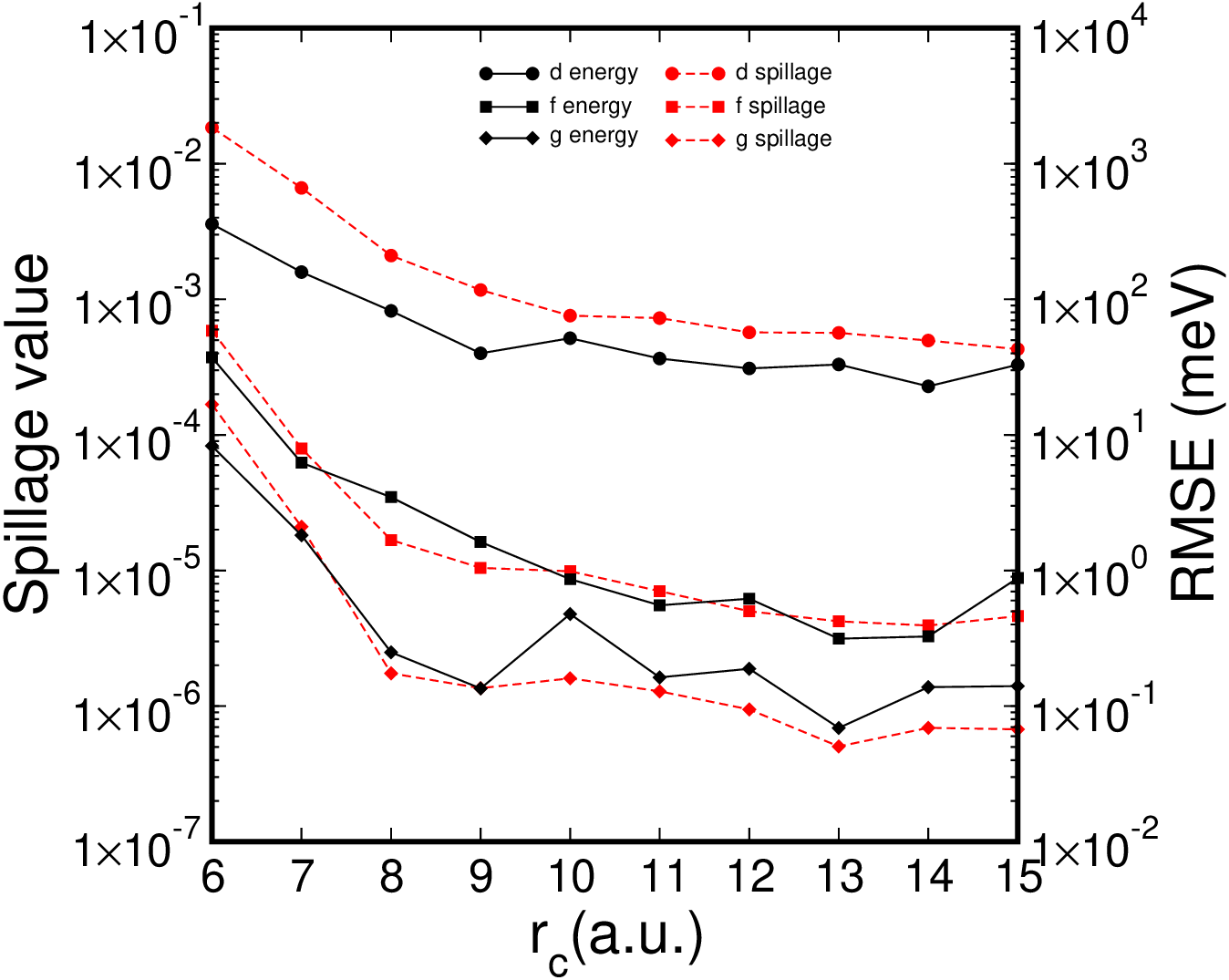}
\caption{(Color online) The spillage values and the averaged RMSE as functions of radial
  cutoff $r_c$ for the basis sets, whose maximal angular momentum are d, f, g,
  respectively.}
\label{fig:Na}
\end{figure}

Figure~\ref{fig:Na} illustrates how the angular momentums and radial cutoff
affect the interpolation accuracy. The black lines
indicate the spillage values while the red lines indicate the averaged RMSE of
the 6 bands.
The radius cutoff changes from 6 a.u. to 15 a.u..
The largest angular momentum of the orbitals is set from $l=2$ (d
orbitals) to $l=4$ (g orbitals). We use 5 radial orbitals (zeta functions)
for each angular momentum, which ensures that the accuracy
improved by multi-zeta orbitals is converged.
We find consistent change between spillage values and the RMSE, 
confirming that the spillage is a good criterion for the
interpolation quality.
The RMSE reduce considerably as $r_c$ increases from 6
a.u. to 9 a.u., while only small improvement is found when $r_c$ is
larger than 9 a.u..
The d orbitals are not sufficient for the high accuracy
interpolation, because the corresponding RMSE are
around 50 to 100 meV even a very large $r_c$ is used. Adding in f
orbitals improve greatly upon d orbitals. For example, for $r_c$=10
a.u., adding f orbitals to the interpolation basis reduces the RMSE
to 0.86 meV.
The g orbitals improve further upon f orbitals, but not as
much as f orbitals to d orbitals.
Since the spillage can be as small as 10$^{-6}$, one can
get highly accurate physical properties related to the wave
functions.

The interpolation results for Si diamond structure and fcc Al are
shown in Fig.~\ref{fig:Si_Na_Al}(b)(c) respectively. For these two
materials, we find g orbitals must be included to get highly accurate
interpolation results. For Si, the lowest 8 bands are interpolated.
The 1s1p1d1f1g orbitals can interpolate the band structures with
RMSE are about 10 meV for all the bands. If the basis
set is doubled, the RMSE for all 8 bands are smaller
than 2 meV, showing extremely good accuracy. When it further 
increases to the level 3 orbitals, the RMSE can be reduced
within 1 meV.
%
For Al,  if the first 1s1p1d1f1g orbitals are used, the RMSE 
can be reduced to 10 meV. When we double the basis to
2s2p2d2f2g, and RMSE of all the bands fall within 3 meV. The RMSE
can be further reduced to less than 1 meV, when
3s3p3d3f3g orbitals are used, also showing remarkable accuracy.

\begin{figure}
\centering
\includegraphics[width=3.in]{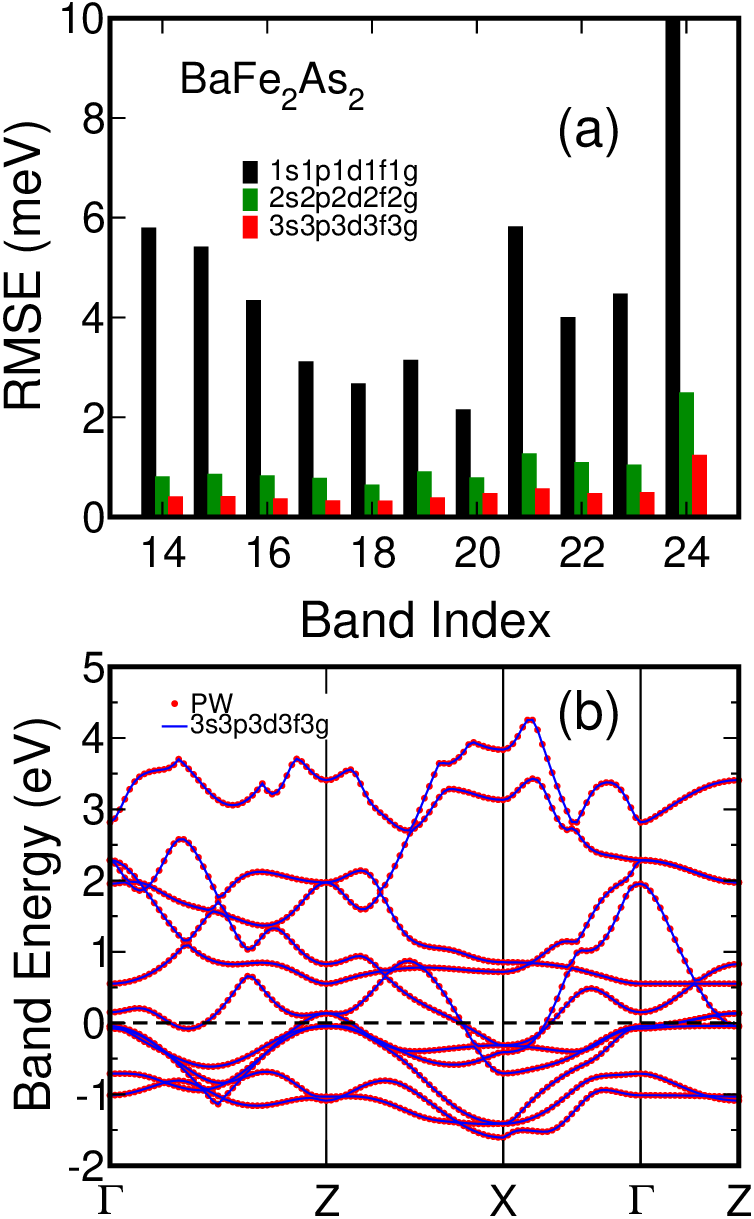}
\caption{ (Color online) (a) The RMSE of interpolated
bands for BaFe$_{2}$As$_{2}$. The black, red, and green columns
represent the RMSE of Level 1, 2, 3 atomic basis sets respectively.
(b) Compare the interpolated band structures (blue lines) using
3s3p3d3f3g basis to the plane wave calculations (red dots).}
\label{fig:BaFe2As2}
\end{figure}

\subsection{BaFe$_{2}$As$_{2}$}

The recent discovery of superconductivity in doped iron arsenide has
attracted great attention. \cite{kamihara08}
Here we show our electronic structure interpolation scheme also works well
for materials like BaFe$_{2}$As$_{2}$ which have complex band structures.
We interpolate the 14th to 24th bands, which span about a 
6 eV energy window around the Fermi level.
%
Figure~\ref{fig:BaFe2As2}(a) depicts the RMSE for each band using
different numbers of numerical orbitals. When using the 1s1p1d1f1g
orbital (25 basis per atom), the RMSE of all the bands
are under 6 meV except for the last one, which is 15.7 meV. Once the
second radial functions are used, the RMSE of all bands
are reduced to below 1.3 meV except the last one which is 2.48 meV.
When using the third radial functions, all the RMSE are
under 0.6 meV except the last one which is 1.23 meV.

Figure~\ref{fig:BaFe2As2}(b) depicts the band structures of
BaFe$_{2}$As$_{2}$ along the $\Gamma$--Z--X--$\Gamma$--Z line. The
band structures calculated by plane waves are shown in blue dotted
line, whereas the band structure calculated by 3s3p3d3f3g numerical
orbitals are plotted in black solid lines. The
interpolated bands and the full {\it ab initio} calculations
are essentially indistinguishable.


We also test GaAs, fcc Cu and CuZn alloys, and the results are
all similar to the the examples above. For GaAs and fcc Cu,  
the RMSE are less than 10 meV by using the 1s1p1d1f1g orbitals
(25 atomic orbitals per atom) only. 
We find that the 3s3p3d3f orbitals (48 orbitals per atom) can interpolate the 
60 bands around fermi level of CuZn alloys with RMSE $\sim$ 2 meV, 
which are quite similar to the results given in Ref.
\onlinecite{prendergast09}, in which 42
bases per atom were used to get the same accuracy. \cite{prendergast09}

\begin{figure}
\centering
\includegraphics[width=3.2in]{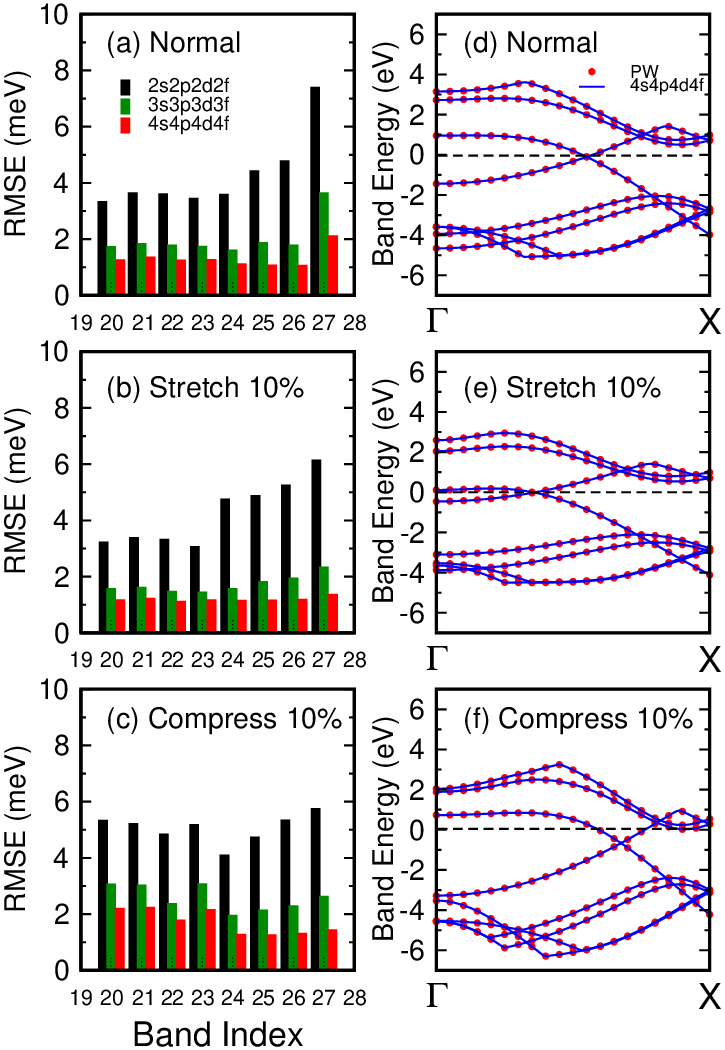}
\caption{(Color online) The RMSE of the interpolated bands for
the (3,3) carbon nanotube of (a) normal, (b) with 10\% stretching, and
(c) with 10\% compression, structures. (d), (e), (f) show
corresponding interpolated band structures (blue lines) using
4s4p4d4f basis compared to the plane wave results (red dots). }
\label{fig:CNT}
\end{figure}

\subsection{TRANSFERBILITY}

Sometimes we need to calculate the physical
properties under certain perturbations, such as strain, defects,
electric field, etc. In these cases, a transferable basis set is
highly desirable. Shirley's Bloch interpolation does not have any
transferbility. Even with a small change of the
system (e.g., a deformation of the unit cell), the Bloch basis needs
to be regenerated. It has been shown that Wanner functions do
not have satisfactory transferbility either, \cite{hierse96,chen09}
because they have too many details. Here we demonstrate that the CGH
orbitals have very good transferbility for the interpolation
purpose.

We take the (3,3) carbon nanotube (CNT) as an example, which is a
metallic armchair nanotube. The system contains 12 atoms. We
interpolate the 20th to 27th bands around the Fermi level.
We use the energy cutoff $E_{c}$=100 Ry.
We first generate the interpolation basis for the CNT at its
equilibrium structure.  We use s, p, d and f orbitals as the
interpolation basis, with the radius cutoff $r_c$=10 a.u. The
RMSE are shown in Fig.~\ref{fig:CNT}(a). The conduction
bands are more difficult to interpolate than valence bands, but
still for Level 2, 3, 4 orbitals, the RMSE are less
than 8 meV, 4 meV and 2 meV respectively. The interpolated band
structures are shown in Fig.~\ref{fig:CNT}(d) in black solid lines
compared to the plane wave results shown in the red dotted lines.
The two band structures are hardly distinguishable.

Using the {\it same} interpolation basis, we calculate the band
structures for (3,3) CNTs which are stretched 10\% or compressed
10\% along the tube and their RMSE. The RMSE
are given in Fig.~\ref{fig:CNT}(b)(c), and the corresponding band
structures are shown in Fig.~\ref{fig:CNT}(e)(f) respectively. As we
see, the RMSE almost do not change with respect to the
CNT stretching and compressing, even though the band
structures themselves change dramatically. These
results demonstrate that the CGH orbitals have excellent
transferbility for the interpolation purpose, which is very
important when perturbation calculations are needed.

\section{Conclusion}
\label{sec:conclusion}

We have presented an efficient scheme for accurate electronic
structure interpolations based on the systematically improvable
optimized atomic orbitals. The current scheme has several advantages
over the existing interpolation schemes. We
find usually 16 -- 25 orbitals per atom can give accuracy about 10
meV compared to the full {\it ab initio} calculations,
similar to that of the revised Shirley's scheme. However, unlike the
Shirley's method, in which a large number of three-dimensional wave
functions have to be stored in the calculations, the present scheme
only needs to store the one-dimensional radial functions. The atomic
orbitals are highly localized, therefore the scheme has many good
features as the interpolation schemes based on maximally localized
Wannier functions. Even though the number of atomic orbitals is
greater than that of Wannier functions, generally the atomic orbitals
are more localized than Wannier functions, and therefore have much less
neighboring atoms. The scheme is easy to implement and robust,
working equally well for metallic systems and systems with complex
band structures. Furthermore, the atomic orbitals have much better
transferability than the Shirley's basis and Wannier functions,
which is very useful for the perturbation calculations.

\acknowledgements

LH acknowledges support from the Chinese National
Fundamental Research Program 2011CB921200, ``Hundreds of Talents'' program
from the Chinese Academy of Sciences
and National Natural Science Funds for Distinguished Young Scholars.


%

\end{document}